\documentclass[12pt]{iopart}

\usepackage{iopams}  

\usepackage{graphicx}
\usepackage{dcolumn}
\usepackage{bm}
\usepackage{epstopdf}
\usepackage{mathrsfs}

\def\bs{\boldsymbol}
\def\del{\partial}

\def\p{{\boldsymbol p}}

\def\bkappa{{\boldsymbol \kappa}}
\def\bbkappa{\bar{\boldsymbol \kappa}}

\def\qqb{{q\bar q}}
\def\sM{\textrm{\scriptsize med}}
\newcommand{\beq}{\begin{eqnarray}}
\newcommand{\eeq}{\end{eqnarray}}
\newcommand{\be}{\begin{eqnarray*}}
\newcommand{\ee}{\end{eqnarray*}}

\newcommand{\nn}{\nonumber\\ }

\begin{document}

\title[Onset of  color decoherence for soft gluon radiation in medium ]{Onset of  color decoherence for soft gluon radiation in medium}

\author{Y Mehtar-Tani$^1$, C A Salgado$^1$ and K Tywoniuk$^2$}

\address{$^1$ Departamento de F\'isica de Part\'iculas and IGFAE,
Universidade de Santiago de Compostela, E-15782 Santiago de Compostela, 
Galicia-Spain}
\address{$^2$ Department of Astronomy and Theoretical Physics, 
Lund University, \\
SE-223 62 Lund, Sweden}

\ead{\mailto{mehtar@fpaxp1.usc.es},\mailto{carlos.salgado@usc.es},\mailto{konrad.tywoniuk@thep.lu.se}}
\begin{abstract}
We report on recent studies of the phenomenon of color decoherence in jets in QCD media. The effect is most clearly observed in the radiation pattern of a quark-antiquark antenna, created in the same quantum state, traversing a dense color deconfined plasma. Multiple scattering with the medium color charges gradually destroys the coherence of the antenna. In the limit of opaque media this ultimately leads to independent radiation off the antenna constituents. Accordingly, radiation off the total charge vanishes implying a memory loss effect induced by the medium.
\end{abstract}

\pacs{12.38.-t,24.85.+p,25.75.-q}

\section{Introduction}

Color coherence is a key feature of jet physics in perturbative QCD. This phenomenon is a direct manifestaion of current conservation at the level of quarks and gluons. Remarkably, these features survive the hadronization process and embodies some of the most striking evidences of color degrees of freedom in collider experiments. As an example, the depletion of soft particles inside a jet is realized in the parton shower by the angular ordering of successive parton branchings. This ordering is indeed a direct consequence of the coherent nature of the jet fragmentation \cite{bas83,Dokshitzer:1978hw,Mueller:1981ex,Ermolaev:1981cm,Dokshitzer:1991wu}. 

Motivated by the new measurments of jets in heavy-ion collisions at RHIC \cite{Putschke:2008wn} and LHC  \cite{Aad:2010bu,Chatrchyan:2011sx}, recently we started to investigate how QCD coherence is altered in the presence of a hot and dense quark-gluon plasma (QGP) \cite{MehtarTani:2010ma,MehtarTani:2011tz,MehtarTani:2011jw}. Before addressing the full-scale problem of medium-induced modifications of jet fragmentation, we started out by studying modification of the radiation pattern of a quark-antiquark antenna traversing a hot and dense medium in order to understand how the QGP alters color coherence of such a system. Unexpectedly, we found a strict geometrical separation between in-vacuum and medium-induced gluon radiation, determined by the antenna opening angle. Additionally, a soft logarithmic divergence remains for the medium-induced spectrum which points to the possibility of resumming multiple gluon branchings in the cascade. In the case of an opaque medium, dictated by unitarity, a simple and intuitive physical picture arises: the interaction with the QGP leads to the gradual decoherence of the pair yielding a universal radiation pattern which does not depend on the initial color configuration of the antenna. These results are in qualitative agreement with the recent data from the LHC and provide a starting point for further studies on in-medium jet modifications.

\section{Antenna radiation pattern to leading log accuracy}
\begin{figure}[t!]
\centering
\includegraphics[width=0.6\textwidth]{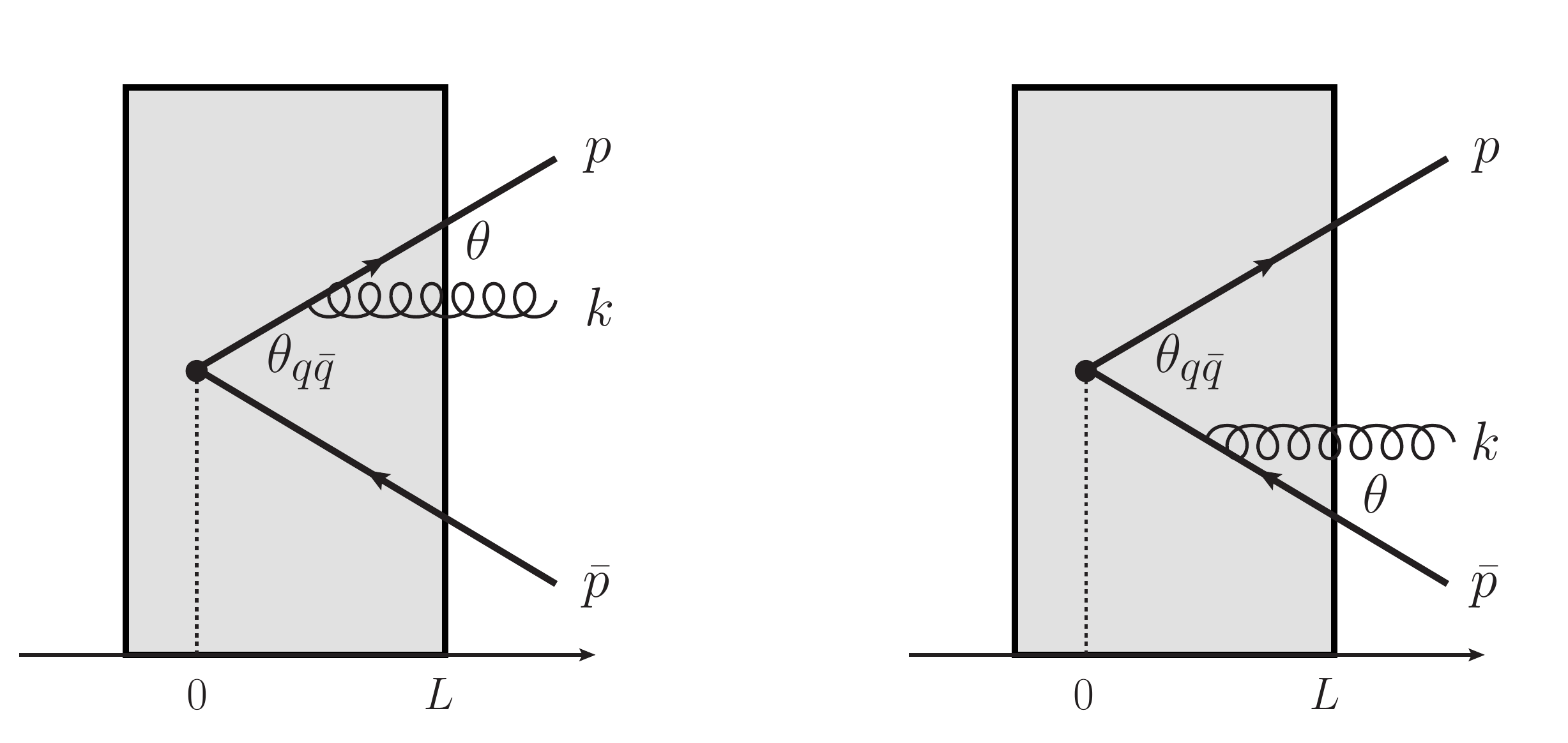}
\caption{Diagrammatic representation of the the gluon radiation amplitude of an energetic quark-antiquark antenna produced inside a QCD medium. }
\label{fig:amplitude}
\end{figure}

We proceed in the framework of the classical Yang-Mills equations, $[D_\mu,F^{\mu\nu}] = J^\nu$, with $D_\mu\equiv \del_\mu-ig A_\mu$ and $F_{\mu\nu}\equiv \del_\mu A_\nu-\del_\nu A_\mu-ig[A_\mu,A_\nu]$. The covariantly conserved current, i.e., $[D_\mu,J^\mu]=0$,  describes the projectiles, in our case a quark and an antiquark  of momentum $p$ and $\bar p$ respectively, born at $x_0=0$ inside the medium of length $L$. Furthermore, we choose the Ligh-cone gauge (LCG) $A^+=0$, where only the transverse polarization contribute to the cross-section. Then the gauge field is related to the gluon radiation amplitude (for a gluon of momentum $k\equiv(\omega,\vec k)$) trough the reduction formula, ${\cal M}_\lambda ^{a}({\vec k})=\lim\limits_{k^2\to 0 } -k^2A^{a}_\mu(k)  \epsilon^\mu_\lambda({\vec k})$. The current of the energetic quark (and similarly for the antiquark), of color charge $Q_q$ ,  gets simply color rotated when passing through the medium,$J^{\mu}_{q}(x)=gU_p(x^+,0)\, \delta^{(3)}(\vec x-\vec p/E\, t)\,Q_q$, where the Wilson line along the trajectory of the quark in the adjoint representation, $U_p$, is given by
\beq
U_p(x^+,0)={\cal P}_+\exp\left[ig\int _0^{x^+} dz^+ T\cdot A_\sM^- \left(z^+,\frac{\p }{p^+}z^+ \right)\right] ,
\eeq
where $A_\sM$ represents the classical medium background-field.
In the strictly soft limit, i.e., $\omega\to 0$,  the current reduces  in Fourier space to
\beq\label{eq:current-sol-soft}
J^{\mu,a}_{q}(k)& =&- ig\frac{p^\mu}{p\cdot k }U_p^{ab}(L,0)\, Q_q^b \,.
\eeq
From the CYM equations, the transverse component of the linear medium response reads \cite{meh07}
\beq
\label{eq:field1}
\square A^i=2ig\left[A_\sM^-,\del^+ A^i\right] -\frac{\del^i}{\del^+}J^++J^i \,.
\eeq
The first term in the r.h.s. of Eq. (\ref{eq:field1}) contains gluon rescattering with the medium, which screens the soft divergence \cite{MehtarTani:2010ma}.
The remainder corresponds to gluon bremsstrahlung where only the quark rescatters and is infrared divergent, see Eq.~(\ref{eq:current-sol-soft}). Keeping only the bremsstrahlung contribution (for finite gluon energy corrections see recent works \cite{MehtarTani:2011jw,CasalderreySolana:2011rz}), the amplitude for soft gluon emission off the quark and antiquark reads (cf. Fig.~(\ref{fig:amplitude}))
\beq
\label{eq:amp-soft1}
{\cal M}_\lambda^{a}(k) = 
-ig\left[\frac{\bkappa\cdot{\bs \epsilon}_\lambda}{x\ (p\cdot k) }U^{ab}_p(L,0)\ Q_q^b+\frac{\bbkappa\cdot{\bs \epsilon}_\lambda}{\bar x\ (\bar p\cdot k )}U^{ab}_{\bar p}(L,0)\ Q_{\bar q}^b\right] \,,
\eeq
where we define the gluon transverse momentum relatively to the quark and the antiquark directions respectively,  $\kappa^i \equiv k^i - x\, p^i$ and $\bar \kappa^i \equiv k^i - \bar x\, \bar p^i$ ($i=1,2$), along with the momentum fractions $x\equiv k^+/p^+\approx \omega/E$ and $\bar x \equiv k^+/\bar p^+\approx \omega/\bar E$.
Putting $U^{ab}=\delta^{ab}$, i.e., in the absence of the medium, the antenna radiation pattern in vacuum is recovered. Squaring the amplitude and summing over the polarization vectors we get the gluon radiation spectrum at leading log accuracy, 
\beq
\label{eq:specoctet}
(2\pi)^2\,\omega\frac{dN_{g^\ast}}{d^3k}& =& \frac{\alpha_s}{\omega^2}\Big\{ C_F\left[ \mathcal{R}_q + \mathcal{R}_{\bar q} -2(1-\Delta_\sM)\,  \mathcal{J} \right]\nn
&&  + C_A(1-\Delta_\sM)\, \mathcal{J} \Big\} \,,
\eeq
where $\mathcal{R}_q = 2/(n_q\cdot n)$, and analogously for the antiquark, correspond to the independent radiation probabilities in vacuum, and $\mathcal{J} = \bkappa\cdot \bbkappa/[\omega^2(n_q\cdot n)(n_{\bar q}\cdot n)]$ corresponds to the interference between the quark and the antiquark radiation in vacuum too, where $n^\mu_q = p^\mu/E$ and $n^\mu = k^\mu/\omega$. In Eq.~(\ref{eq:specoctet}). The first term, proportional to $C_F$ corresponds to the emission off the quark or the antiquark and contains two contributions: vacuum one, confined inside the pair at angles smaller than the antenna opening angle,  $\theta<\theta_\qqb$, due to destructive interferences at large angles ; the medium induced part, proportional to $\Delta_\sM$, occurs at large angles and is suppressed inside the antenna. The second one describes large angle emissions by the total charge of the pair, i.e., $C_A$ for gluon, and vanishes in the case of a virtual photon. 
The spectrum, given by Eq.~(\ref{eq:specoctet}), has a simple form and offers an intuitive physical picture: Interestingly enough, the medium parameters are fully contained in a multiplicative factor, 
\beq
\label{eq:Delta}
\Delta_\sM=1-\frac{1}{N_c^2-1}\langle {\bf Tr}\,U_p(L,0) U_{\bar p}^\dag(L,0) \rangle \,,
\eeq
while the functional shape is vacuum-like. In the dilute limit, $\Delta_\sM\to0$, we recover the pure vacuum spectrum, $dN\to dN^\textrm{\scriptsize vac}$. With increasing density, the decoherence rate is controlled by the parameter $\Delta_\sM$. In the limit of a completely opaque system, $\Delta_\sM$ is bounded by unitarity so that $\Delta_\sM\to 1$. Then the interferences are completely washed out and the soft emissions in the presence of a medium reduces to independent radiation off the quark and antiquark, as if they were radiating in the vacuum. This is what we call {\it total decoherence} of the spectrum. This implies a memory loss effect in the medium, so that
\beq
dN_{g^\ast}\Big|_\textrm{\scriptsize opaque} = dN_{\gamma^\ast}\Big|_\textrm{\scriptsize opaque} \,,
\eeq
i.e., the antenna radiation is independent of the total color charge.

This result can be summarized as follows: two charges born in the same quantum state, and therefore expected to radiate coherently, will become two independent emitters when traversing an opaque medium, in effect loosing memory of their origin. As a consequence, angular ordering in the parton shower should be partially (or completely) destroyed. 

While the complete description of jet fragmentation in the presence of a medium still is missing, our results provide a starting point for further studies and suggests that the probabilistic nature of the fragmentation process for soft radiation survives. This calls for a reformulation of Monte-Carlo codes which aim at describing high-p$_T$ physics in heavy-ion collisions.

In summary, we have computed the antenna spectrum to leading logarithmic accuracy in the presence of a QCD medium which interpolates between the dilute and dense limiting cases and demonstrates the gradual onset of decoherence. The modification of the interference properties of radiation off coherent quantum systems we have found are genuine properties of the strong interaction and will therefore also affect other types of QCD coherence, such as the drag-effect and color flow in hadronic collisions \cite{Dokshitzer:1991wu}.

This work is supported by Ministerio de Ciencia e Innovaci\'on of Spain; by Xunta de Galicia ; by the Spanish Consolider-Ingenio 2010 Programme CPAN; by the European Commission and in part by the Swedish Research Council (contract number 621-2010-3326). CAS is a Ram\'on y Cajal researcher.

\end{document}